\documentclass[prb,twocolumn,showpacs]{revtex4}
\usepackage{color,soul}
\usepackage{amsmath}
\usepackage{dcolumn}
\usepackage{graphicx}
\usepackage{bm}
\usepackage{amssymb}
\usepackage{subfigure}
\begin{document}

\title{On the origin of the possible chiral spin liquid state of the triangular lattice Hubbard model}
\author{Qiu Zhang and Tao Li}
\affiliation{Department of Physics, Renmin University of China, Beijing 100872, P.R.China}

\begin{abstract}
Recent numerical simulations find a possible chiral spin liquid state in the intermediate coupling regime of the triangular lattice Hubbard model. Here we provide a simple picture for its origin in terms of a Bosonic RVB description. More specifically, we show that such a chiral spin liquid state can be understood as a quantum disordered tetrahedral spin state stabilized by the four spin ring exchange coupling, which suppresses the order-by-disorder effect toward a stripy spin state. Such a chiral spin liquid state features a spin Berry phase of $\frac{\pi}{2}$ per triangle. However, we show that the topological property of such a state is totally missed in the Schwinger Boson mean field description as a result of the lack of Boson rigidity caused by the no double occupancy constraint.
\end{abstract}

\maketitle

The search of spin liquid state in the intermediate coupling regime of the triangular lattice Hubbard model(TLHM) is triggered by the discovery of a possible $U(1)$ spin liquid state with a large spinon Fermi surface in organic salt compounds\cite{Kanoda1,Kanoda2,Matsuda,Zhou}. Such a state can be understood roughly as the descendant of a metallic state near a Mott transition, in which electron correlation has already opened a charge gap while leaving the electron Fermi surface intact. An insulator with a large Fermi surface is exotic in the sense that the gapless quasiparticles on the Fermi surface should carry only the spin but not the charge quantum number of an electron and is intrinsically fractionalized. Variational studies find that when the multi-spin exchange coupling is strong enough, as is the case near a Mott transition, a $U(1)$ spin liquid state with a large spinon Fermi surface is the best variational ground state of a quantum antiferromagnet defined on the triangular lattice\cite{Motrunich,Ronny}.

However, the fate of such a novel spin liquid state depends crucially on the size of the multi-spin ring exchange coupling that one can reach before the collapse of the charge gap. Recently, DMRG simulations indicate that the critical value for charge gap opening($U_{c1}/t$) is significantly larger than earlier estimation, making the $U(1)$ spin liquid state with a large spinon Fermi surface less likely to occur in the pure TLHM. In the intermediate coupling regime between $U_{c1}/t$ and the transition point toward the 120$^{\circ}$ ordered phase($U_{c2}/t\approx 11$), evidences in support of a chiral spin liquid breaking both the time reversal and parity symmetry are found numerically\cite{Szasz,Chen,Cookmeyer,Wietek}.

In this paper, we provide a simple picture for the origin of such a chiral spin liquid state in terms of the Bosonic resonating valence bond(RVB) theory. We show that the observed chiral spin liquid state can be understood as a quantum disordered tetrahedral spin state, which features a spin Berry phase of $\frac{\pi}{2}$ per triangle of the triangular lattice. The realization of the chiral spin liquid state by quantum disordering such an non-coplanar state has also been proposed in other models previously\cite{Paramekanti}. We show that the four spin ring exchange coupling can suppress the order-by-disorder effect toward a collinear stripy spin state on the triangular lattice and favors a non-coplanar tetrahedral spin state, which has equal length but orthogonal spin order parameter at all the three degenerate stripy ordering wave vectors. 

To illustrate such a picture, we have performed a Schwinger Boson mean field calculation on the $J_{1}-J_{2}-K_{4}$ model on the triangular lattice. The model reads
\begin{eqnarray}
H&=&J_{1}\sum_{<i,j>}\mathbf{S}_{i}\ \cdot \mathbf{S}_{j}+J_{2}\sum_{<<i,j>>}\mathbf{S}_{i}\ \cdot \mathbf{S}_{j}\nonumber\\
&+&K_{4}\sum_{[i,j,k,l]}(P_{ijkl}+P_{ilkj}),
\end{eqnarray}
in which $J_{1}$, $J_{2}$ and $K_{4}$ denote the Heisenberg exchange coupling on the first neighboring bonds, the second neighboring bonds and the four spin ring exchange coupling around the elementary rhombi of the triangular lattice. Such an effective spin model can be realized in the strong coupling expansion of the TLHM\cite{Yang} and had been studied by many researchers\cite{LiMing,Motrunich,Ronny1,Xu,Seki,Cookmeyer}. We find that a chiral spin state prevails in a large portion of the phase diagram of this model. We note that the $K_{4}$ term can also be expressed in terms of the spin operator as 
\begin{eqnarray}
P_{ijkl}+P_{ilkj}=4[(\mathbf{S}_{i}\ \cdot \mathbf{S}_{j})(\mathbf{S}_{k}\ \cdot \mathbf{S}_{l})+(\mathbf{S}_{i}\ \cdot \mathbf{S}_{l})(\mathbf{S}_{j}\ \cdot \mathbf{S}_{k})\nonumber\\
-(\mathbf{S}_{i}\ \cdot \mathbf{S}_{k})(\mathbf{S}_{j}\ \cdot \mathbf{S}_{l})]+\frac{1}{2}(\mathbf{S}_{i}+\mathbf{S}_{j}+\mathbf{S}_{k}+\mathbf{S}_{l})^{2}-\frac{5}{4}.\nonumber\\
\end{eqnarray}
Using such an  identity,  one find that the model parameterization used here is related to that used in Ref.[\onlinecite{Cookmeyer}] by $J_{1}=J_{1}+5K_{4}$, $J_{2}=J_{2}+K_{4}$, $J_{4}=4K_{4}$.
  
We start from a Fermionic RVB analysis on the origin of the possible chiral spin liquid state in the limiting case of $J_{2}=0$. When $K_{4}=0$, the model reduces to the triangular lattice antiferromagnetic Heisenberg model(TLHAF), which prefers a 120$^{\circ}$ coplanar order in its ground state. In the Fermioinc RVB formulation, it is well known that such an ordering tendency is captured by a Dirac spin liquid state with a $\pi$-flux enclosed in each plaquette of the triangular lattice\cite{Yunoki,Becca2019,Tao20201, Zhangqiu}.  The mean field ansatz for such a Dirac spin liquid state is given by
\begin{equation}
H_{MF}=-\sum_{<i,j>,\alpha}\chi_{i,j} f^{\dagger}_{i,\alpha}f_{j,\alpha},
\end{equation}
in which $\chi_{i,j}$ is a hopping integral of the Fermionic spinon. In the Dirac spin liquid state, the consecutive product of $\chi_{i,j}$ around each plaquette of the triangular lattice encloses a gauge flux of $\pi$(see Fig.1a for an illustration). On the other hand, when $K_{4}/J_{1}$ is large, the system prefers a $U(1)$ spin liquid state with a large spinon Fermi surface, which is described by a mean field ansatz with zero gauge flux within each plaquette\cite{Motrunich}(see Fig.1c for an illustration). Thus it is reasonable to imagine a spin liquid state with $\frac{\pi}{2}$ gauge flux within each triangle for intermediate value of $K_{4}/J_{1}$(see Fig.1b for an illustration). The mean field ansatz of such a spin liquid state is similar to that of the Dirac spin liquid state in that they enclose the same gauge flux of $\pi$ around each plaquette of the triangular lattice and is similar to that of the $U(1)$ spin liquid state in that the gauge flux distributes uniformly on all triangles. Such an intermediate state is nothing but a chiral spin liquid state and has a nonzero and uniform expectation value for the scalar spin chirality $\chi_{i,j,k}=\langle \mathbf{S}_{i}\cdot(\mathbf{S}_{j}\times\mathbf{S}_{k})\rangle$ on each triangle of the triangular lattice\cite{Laughlin,Wen}.
  
\begin{figure}
\includegraphics[width=8cm]{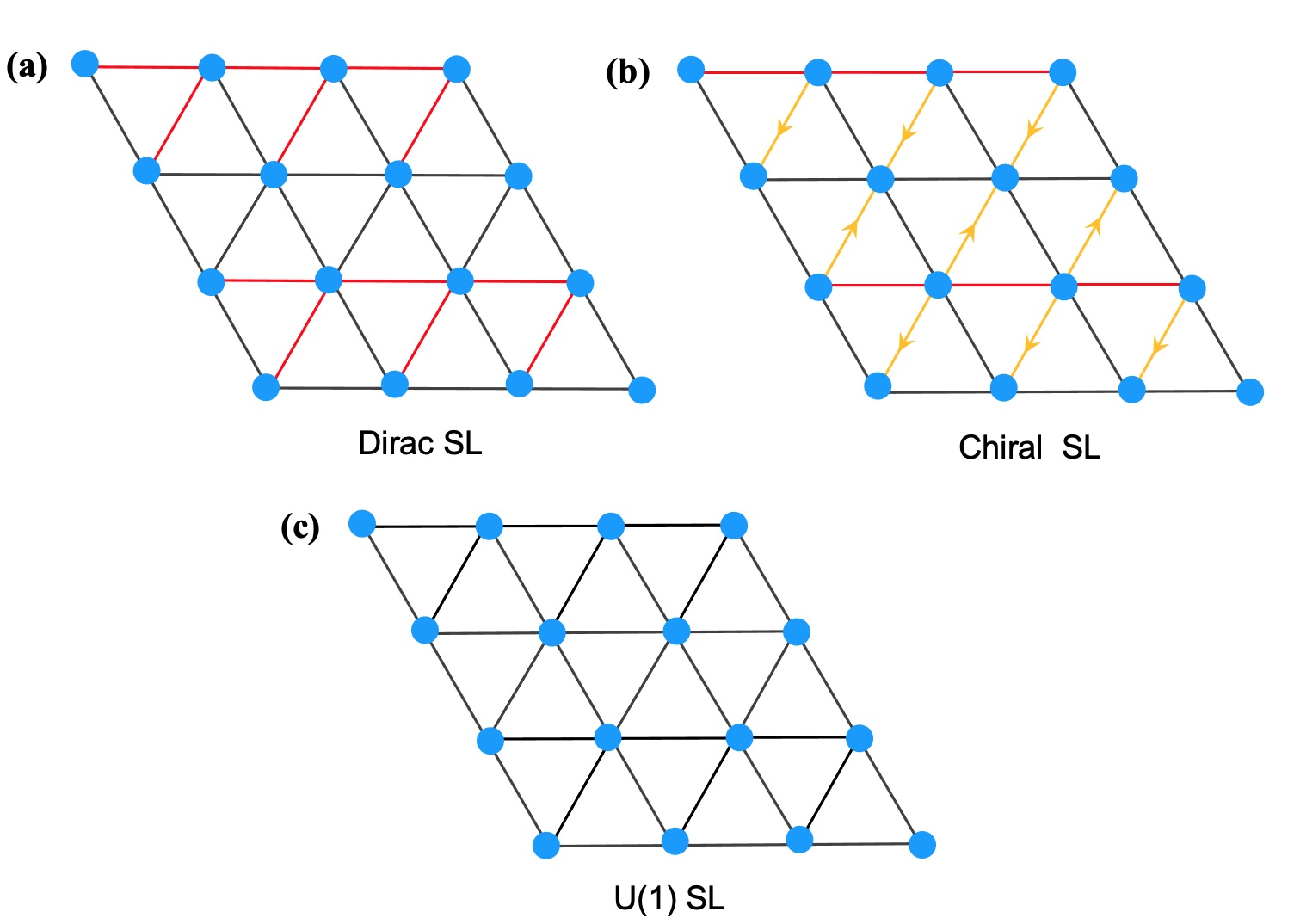}
\caption{Illustration of the Fermionic mean field ansatz for (a)the Dirac spin liquid state, (b)the chiral spin liquid state and (c)the $U(1)$ spin liquid state on the triangular lattice. The hopping integral $\chi_{i,j}$ on the black and red bonds are 1 and -1. The hopping integral on the orange bonds are $i$ along the direction of the arrow and are otherwise $-i$. In both the Dirac spin liquid state and the chiral spin liquid state, the hopping integral $\chi_{i,j}$ encloses a gauge flux of $\pi$ around each plaquette of the triangular lattice. The difference between the two states lies in the fact that the gauge flux distributes uniformly on all triangles in the chiral spin liquid state. Previous studies find that the Dirac spin liquid state is favored at small value of $K_{4}/J_{1}$ and that the $U(1)$ spin liquid state is favored at large value of $K_{4}/J_{1}$. It is thus reasonable to expect the chiral spin liquid state to emerge at intermediate value of $K_{4}/J_{1}$.}
\end{figure}    

A similar, but physically much more transparent understanding on the origin of the chiral spin liquid state can be achieved within the Bosonic RVB formalism, which has a better connection with the semiclassical limit. To begin with, we note that in the chiral spin liquid state, the spinon experiences a Berry phase of $\frac{\pi}{2}$ when going around a triangle. The same spin Berry phase can be generated by a classical spin ordering background with the ordered spins spanning a solid angle of $\pi$ around each triangle. More generally, the uniform distribution of the spin Berry phase of $\frac{\pi}{2}$ within each triangle in the chiral spin liquid state can be generated by a tetrahedral spin ordering pattern as is illustrated in Fig.2b. We thus expect that quantum disordering the tetrahedral ordering pattern by quantum fluctuation will drive the system into the chiral spin liquid state.

\begin{figure}
\includegraphics[width=8cm]{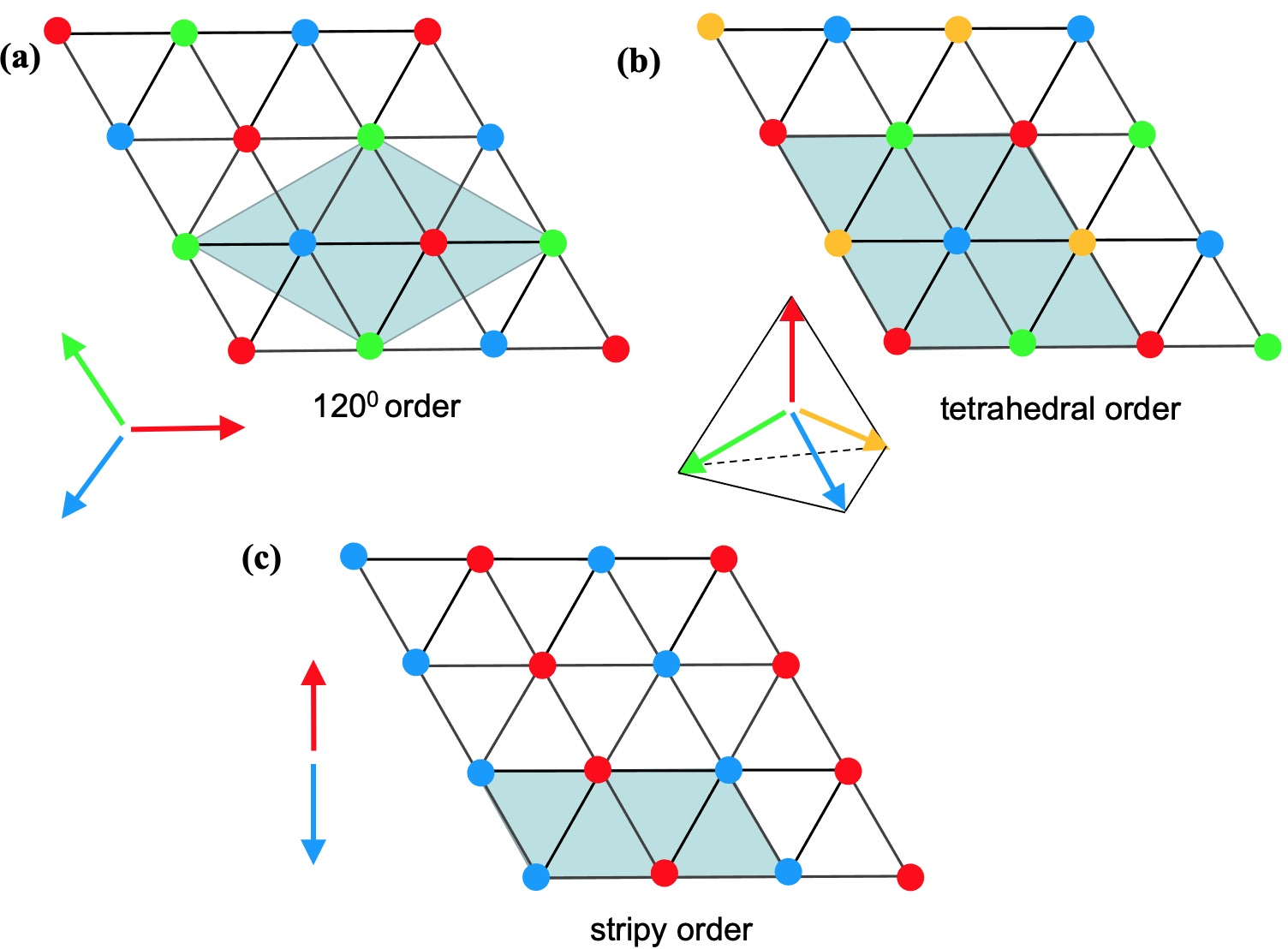}
\caption{Illustration of the three classical spin orders considered in this work. (a)The 120$^{\circ}$ coplanar order. We choose the ordered moment in the three sublattices as $\mathbf{S}_{1}=m(1,0,0)$(red), $\mathbf{S}_{2}=m(-\frac{1}{2},\frac{\sqrt{3}}{2},0)$(green) and $\mathbf{S}_{3}=m(-\frac{1}{2},-\frac{\sqrt{3}}{2},0)$(blue). (b)The non-coplanar tetrahedral order. We choose the ordered moment in the four sublattices as  $\mathbf{S}_{1}=\frac{m}{\sqrt{3}}(1,1,1)$(red), $\mathbf{S}_{2}=\frac{m}{\sqrt{3}}(-1,1,-1)$(green), $\mathbf{S}_{3}=\frac{m}{\sqrt{3}}(-1,-1,1)$(blue) and $\mathbf{S}_{4}=\frac{m}{\sqrt{3}}(1,-1,-1)$(orange). (c)The collinear stripy order. We choose the ordered moment in the two sublattices as  $\mathbf{S}_{1}=m(0,0,1)$(red) and $\mathbf{S}_{2}=m(0,0,-1)$(blue). The gray area in (a), (b) and (c) denotes the magnetic unit cell in the three situations. Note that there are three degenerate stripy ordering patterns on the triangular lattice. Here we only illustrate one of them.}
\end{figure}    

As a non-coplanar state, the development of the tetrahedral spin order requires condensation of spin order parameter on multiple degenerate wave vectors\cite{Martin,TaoLi}. On the triangular lattice, there is a natural way to meet such a requirement. In fact, it is well known that for the $J_{1}-J_{2}$ model on the triangular lattice, the Fourier transform of the exchange coupling has three degenerate minimum corresponding to the stripy order along the three different directions when $J_{2}>\frac{J_{1}}{8}$(see Fig. 2c for an illustration). The tetrahedral spin state is generated when mutually orthogonal spin order parameters of the same magnitude are condensed at all the three degenerate stripy ordering wave vectors. However, usually the collinear order with spin condensation on a single stripy wave vector is chosen as a result of the quantum or thermal order-by-disorder effect. We show that such an effect can be efficiently suppressed by the four spin ring exchange coupling. More specifically, the expectation value of the ring exchange terms in the classical ground state of the 120$^{\circ}$, the tetrahedral and the stripy ordered phase are $-\frac{21}{8}K_{4}$, $-\frac{11}{3}K_{4}$ and $-3K_{4}$ per site respectively. Thus we expect that the chiral spin state to be stabilized by the $K_{4}$ term.

Now we provide a Bosonic RVB description of such a chiral spin state, in which the spin operator is written in terms of the Schwinger Boson operator as
\begin{equation}
\mathbf{S}_{i}=\frac{1}{2}\sum_{\alpha,\beta}b^{\dagger}_{i,\alpha}\bm{\sigma}_{\alpha,\beta}b_{i,\beta}.
\end{equation}
Here $\alpha,\beta=\uparrow,\downarrow$ is the spin index of the Bosonic spinon operator, $\bm{\sigma}$ is the usual Pauli matrix. The spinon operator should satisfy the constraint of 
\begin{equation}
\sum_{\alpha}b^{\dagger}_{i,\alpha}b_{i,\alpha}=1
\end{equation} 
to be a faithful representation of the spin algebra. In the Schwinger Boson formulation, two types of RVB order parameter are introduced to describe the the structure of a spin liquid state. They are the pairing parameter
\begin{equation} 
A_{i,j}=\frac{1}{2}\langle (b_{i,\uparrow}b_{j,\downarrow}-b_{i,\downarrow}b_{j,\uparrow})\rangle\nonumber\\
\end{equation}
describing antiferromagnetic correlation between site $i$ and $j$ and the hopping parameter
\begin{equation}
B_{i,j}=\frac{1}{2}\langle (b^{\dagger}_{i,\uparrow}b_{j,\uparrow}+b^{\dagger}_{i,\downarrow}b_{j,\downarrow})\rangle,
\end{equation}
describing the ferromagnetic correlation between site $i$ and $j$. We note that $A_{i,j}=-A_{j,i}$, $B_{i,j}=B^{*}_{j,i}$.

To write down an RVB mean field ansatz for the chiral spin state we first go to the semiclassical limit of the model, in which we can take the spin operator $\mathbf{S}_{i}$ as classical vectors and the boson operator $b_{i,\alpha}$ as conventional complex numbers. It takes the form of
\begin{eqnarray}
\left(\begin{array}{c}b_{i,\uparrow} \\b_{i,\downarrow}\end{array}\right)&=&\sqrt{m}e^{i\varphi_{i}}\left(\begin{array}{c}\ \cos\frac{\theta_{i}}{2}\ e^{-i\frac{\phi_{i}}{2}} \ \\ \  \sin\frac{\theta_{i}}{2} \ e^{i\frac{\phi_{i}}{2}}\ \end{array}\right).
\end{eqnarray}
Here $\theta_{i}$ and $\phi_{i}$ are the direction angle of the spin vector $\mathbf{S}_{i}$, $m$ is the length of the ordered magnetic moment, $\varphi_{i}$ is an arbitrary gauge phase. The freedom to choose such a gauge phase can be exploited to simplify the form of the mean field ansatz.  Here we choose the ordered moment in the 120$^{\circ}$ ordered phase to lie in the $x-y$ plane. More specifically, we choose
\begin{eqnarray}
\mathbf{S}_{1}&=&m(1,0,0)\nonumber\\
\mathbf{S}_{2}&=&m(-\frac{1}{2},\frac{\sqrt{3}}{2},0)\nonumber\\
\mathbf{S}_{3}&=&m(-\frac{1}{2},-\frac{\sqrt{3}}{2},0).
\end{eqnarray} 
We choose the ordered moment in the stripy ordered phase to be aligned  with the $z$ axis. More specifically, we choose
\begin{eqnarray}
\mathbf{S}_{1}&=&m(0,0,1)\nonumber\\
\mathbf{S}_{2}&=&m(0,0,-1).
\end{eqnarray}
The four ordered moments in the tetrahedral spin state are chosen to be
\begin{eqnarray}
\mathbf{S}_{ 1}&=&\frac{m}{\sqrt{3}}(1,1,1)\nonumber\\
\mathbf{S}_{ 2}&=&\frac{m}{\sqrt{3}}(-1,1,-1)\nonumber\\
\mathbf{S}_{ 3}&=&\frac{m}{\sqrt{3}}(-1,-1,1)\nonumber\\
\mathbf{S}_{ 4}&=&\frac{m}{\sqrt{3}}(1,-1,-1).
\end{eqnarray}

Inserting the value of $b_{i,\alpha}$ we got from the semiclassical analysis into the definition of $A_{i,j}$ and $B_{i,j}$(with a proper choice of the gauge phase $\varphi_{i}$), we can write down the RVB mean field ansatz for all the three spin states. Here we will only present the ansatz for the tetrahedral spin state. The ansatz for the other two states can be found in the Appendix. In the tetrahedral spin state, the RVB parameter on the nearest neighboring bonds are given by
\begin{eqnarray}
A_{1,2}&=&A_{3,4}=\mathcal{A}_{1}e^{i\frac{\pi}{3}}\nonumber\\
A_{1,3}&=&A_{2,4}=\mathcal{A}_{1}e^{i\frac{\pi}{2}}\nonumber\\
A_{1,4}&=&A_{2,3}=\mathcal{A}_{1}e^{i\frac{2\pi}{3}}\nonumber\\
B_{1,2}&=&B_{2,3}=B_{3,4}=-B_{1,4}=\mathcal{B}_{1}\nonumber\\
B_{1,3}&=&-B_{2,4}=-i\mathcal{B}_{1},\nonumber\\
\end{eqnarray}
in which we have abbreviated $A_{i\in1,j\in2}$ as $A_{1,2}$ and so on. $\mathcal{A}_{1}$ and $\mathcal{B}_{1}$ are two real parmeters denoting the amplitude of the RVB parameters on the nearest neighboring bonds. For second neighbors we should replace $\mathcal{A}_{1}$ and $\mathcal{B}_{1}$ with two other real parameters $\mathcal{A}_{2}$ and $\mathcal{B}_{2}$. We note that such a mean field ansatz is gauge equivalent to the ansatz proposed in Ref.[\onlinecite{Messio}](see Appendix for more details). It describes a weakly symmetric spin liquid state which is invariant under space translation, proper space rotation and the combined operation of space inversion and time reversal(but not individually).

The mean field Hamiltonian of the tetrahedral spin state has the form of
\begin{equation}
H_{MF}=\sum_{\mathbf{k}}\Psi^{\dagger}_{\mathbf{k}} \mathbf{H}_{\mathbf{k}}\Psi_{\mathbf{k}},
\end{equation}
in which 
\begin{equation}
\Psi_{\mathbf{k}}=\left(\begin{array}{c} \bm{\phi}_{\mathbf{k}\uparrow} \\ \bm{\phi}^{\dagger}_{\mathbf{-k}\downarrow}\end{array}\right).
\end{equation}
Here 
\begin{equation}
\bm{\phi}_{\mathbf{k}\uparrow}=\left(\begin{array}{c}b_{\mathbf{k},1,\uparrow} \\b_{\mathbf{k},2,\uparrow} \\b_{\mathbf{k},3,\uparrow} \\b_{\mathbf{k},4,\uparrow} \ \end{array}\right); \ \bm{\phi}_{\mathbf{-k}\downarrow}=\left(\begin{array}{c}b^{\dagger}_{\mathbf{-k},1,\downarrow} \\b^{\dagger}_{\mathbf{-k},2,\downarrow} \\b^{\dagger}_{\mathbf{-k},3,\downarrow} \\b^{\dagger}_{\mathbf{-k},4,\downarrow} \ \end{array}\right)
\end{equation}
denote the spinon operator on the four sublattices.
$\mathbf{H}_{\mathbf{k}}$ is a $8\times8$ matrix and is given by
\begin{equation}
\mathbf{H}_{\mathbf{k}}=\left(\begin{array}{cc}\lambda+M_{1} & M_{3} \\M^{\dagger}_{3} & \lambda+M_{2}\end{array}\right),
\end{equation}
in which $\lambda$ is a Boson chemical potential introduced to enforce the single occupancy constraint Eq.5 at the average level. The $4\times4$ submatrices $M_{1}$, $M_{2}$ and $M_{3}$ satisfy 
\begin{eqnarray}
M_{1}&=&M^{\dagger}_{1}\nonumber\\
M_{2}(\mathbf{k})&=&M^{T}_{1}(-\mathbf{k})\nonumber\\
M_{3}(\mathbf{k})&=&-M^{T}_{3}(-\mathbf{k}).
\end{eqnarray} 
The matrix element of $M_{1}$ and $M_{3}$ can be found in the Appendix.
 
The mean field Hamiltonian can be diagonalized by the following para-unitary transformation
\begin{equation}
\Psi_{\mathbf{k}}=U_{\mathbf{k}}\Phi_{\mathbf{k}},
\end{equation}
in which the matrix $U_{\mathbf{k}}$ satisfy the equation
\begin{eqnarray}
U_{\mathbf{k}}\ \Sigma \ U^{\dagger}_{\mathbf{k}}&=&\Sigma \nonumber\\
U^{\dagger}_{\mathbf{k}}\ \mathbf{H}_{\mathbf{k}}\ U_{\mathbf{k}}&=&\Lambda_{\mathbf{k}}.
\end{eqnarray}
Here
\begin{equation}
\Sigma=\left(\begin{array}{cc}\mathbf{I} & 0 \\0 & -\mathbf{I}\end{array}\right);\ \Lambda_{\mathbf{k}}=\left(\begin{array}{cc}\bm{\epsilon}_{\mathbf{k}} & 0 \\0 & \bm{\epsilon}_{\mathbf{k}}\end{array}\right),
\end{equation}
in which $\mathbf{I}$ a $4\times4$ identity matrix, $\bm{\epsilon}_{\mathbf{k}}$ is a $4\times4$ diagonal matrix with diagonal element $\varepsilon^{n}_{\mathbf{k}}$ for $n=1,...,4$. The chemical potential $\lambda$ should be determined by the equation
\begin{eqnarray}
4&=&\frac{1}{2N}\sum_{\mathbf{k}}\langle \Psi^{\dagger}_{\mathbf{k}} \  \Psi_{\mathbf{k}}\rangle\nonumber\\
&=&\frac{1}{2N}\sum_{\mathbf{k}}\sum^{8}_{n=5} \left[ U^{\dagger}_{\mathbf{k}} U_{\mathbf{k}}\right]_{n,n}.
\end{eqnarray}

After solving the mean field equation we can calculate the variational energy of the $J_{1}-J_{2}-K_{4}$ model by applying the Wick decomposition of the model Hamiltonian in the mean field ground state. The variational energy appears as a function of the four real parameter $\mathcal{A}_{1}$, $\mathcal{A}_{2}$, $\mathcal{B}_{1}$ and $\mathcal{B}_{2}$ in the tetrahedral spin state and can be optimized by the simulated annealing method. The same procedure can be performed on the 120$^{\circ}$ spin state and the stripy spin state. We then compare the variational energies generated from these three spin states and determine the mean field phase diagram of the system, the result of which is presented in Fig.3.

\begin{figure}
\includegraphics[width=7cm]{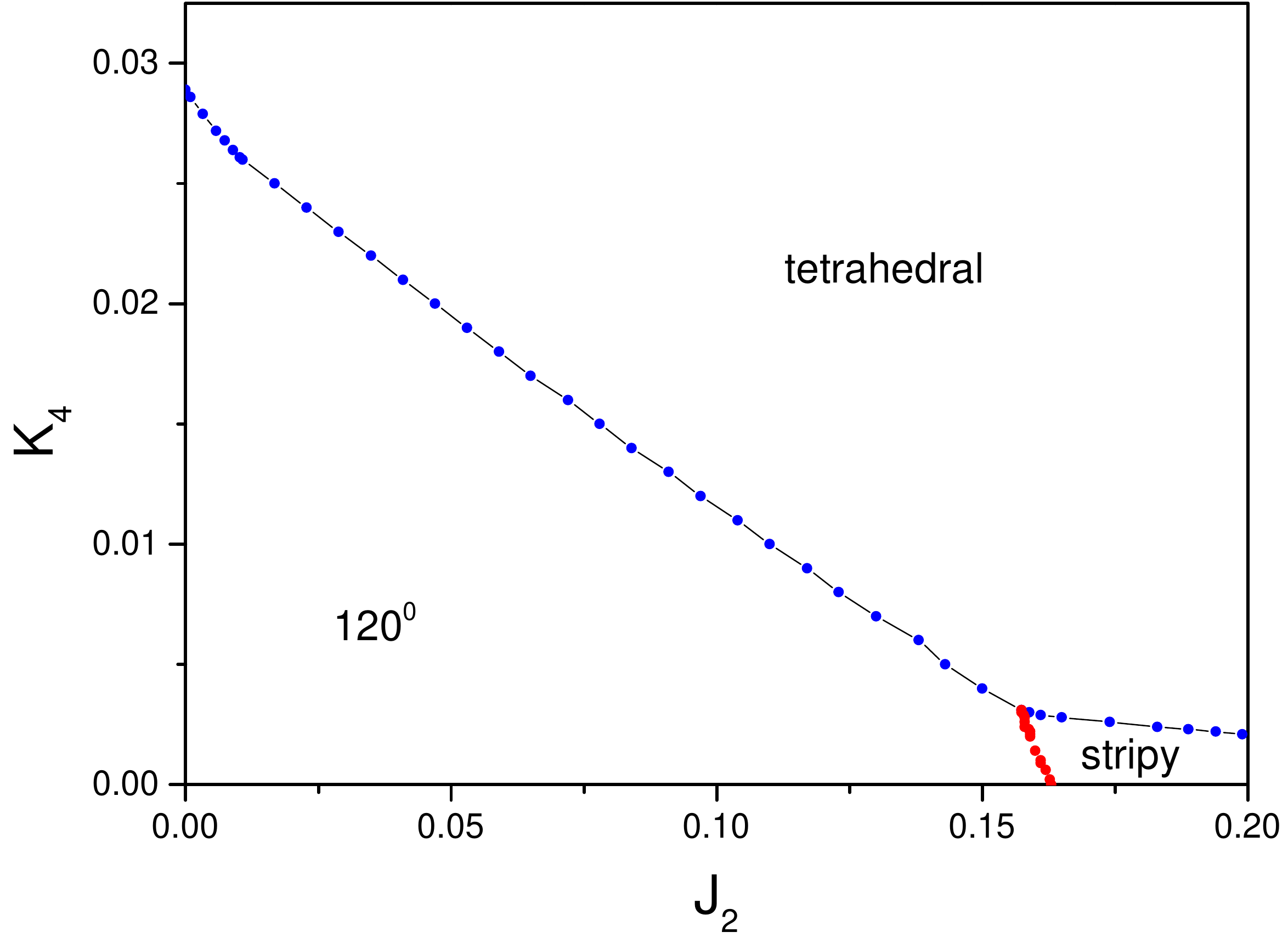}
\caption{The phase diagram of the $J_{1}-J_{2}-K_{4}$ model on the triangular lattice determined by simulated annealing optimization of the Schwinger Boson mean field ground state energy for the 120$^{\circ}$, the tetrahedral and the stripy spin state . The nearest neighboring exchange coupling $J_{1}$ is set as the unit of energy. The phase boundaries in the phase diagram all correspond to first order transition.}
\end{figure}  

We find that the mean field phase digram is qualitatively similar to the classical phase diagram. The tetrahedral spin state indeed prevails over the stripy spin state and the 120$^{\circ}$ spin state in a large portion of the phase digram. We note that for small $J_{2}$ a Zig-Zag spin state will take over the tetrahedral spin state at large $K_{4}$. However, since our main focus is the chiral spin state, we leave the competition of such a state with the tetrahedral spin state at small $J_{2}$ and large $K_{4}$ to future study. From this phase digram, we see that the order-by-disorder effect favoring the stripy spin state can indeed be suppressed by the ring exchange coupling. We also find that the critical value for the appearance of the tetrahedral spin state at $J_{2}=0$ is about $K_{4}\approx0.03$, which is just the order of magnitude expected from the strong coupling expansion of the TLHM around $U/t=10$. 

We note that according to mean field solution the Bosoinc spinon is gapless in the thermodynamic limit throughout the phase diagram, which implies spinon condensation and the spontaneous breaking of spin rotational symmetry. However, we note that the Schwinger Boson mean field theory overestimates the tendency of spinon condensation as a result of the lack of Boson rigidity in the mean field description. For example, it is well known that the mean field theory overestimates the local spin sum rule $\mathbf{S}^{2}=\frac{3}{4}$ by a factor of $\frac{3}{2}$ if we insist on enforcing the Boson density constraint of $\langle (b^{\dagger}_{i,\uparrow}b_{i,\uparrow}+b^{\dagger}_{i,\uparrow}b_{i,\uparrow})\rangle=1$ at the mean field level. When quantum fluctuation caused by such Boson rigidity is taken into account, one expect that the 
spinon condensation will be suppressed and the spin rotational symmetry may be recovered. However, the chiral order which breaks the discrete time reversal symmetry is expected to be more robust and will survive the quantum fluctuation in some region of the phase diagram. 

We also note that the lack of Boson rigidity in the Schwinger Boson mean field theory is fatal for a correct description of the topological property of the system(we note that the lack of Boson rigidity in the Schwinger Boson mean field theory is also fatal for a correct description of the dynamical property of the system\cite{Zhangqiu}). In the Fermionic RVB picture, the chiral spin liquid state is described by a Gutzwiller projected topological insulating state with a nonzero Chern number of $\pm1$. Gapless chiral edge state and quantized thermal Hall response is thus expected from such a spin liquid state, even at the mean field level. However, in the Bosonic RVB picture, the mean field ground state of such a chiral spin state is topological trivial. This can be easily seen from the fact that the state can be deformed continuously(without closing the spinon gap in the process) into the trivial Boson vacuum state by sending the Boson chemical potential to infinity.  It is thus quite interesting to know if the Fermionic and the Bosonic description of such a chiral spin liquid state are indeed equivalent with each other. From the above discussion, we know that such an equivalence can be established only after we enforce the no double occupancy constraint explicitly. We leave such a study to future works.

Lastly, we note that according to the picture presented in this work, the chiral spin liquid state should be understood as the descendent of the tetrahedral ordered spin state, rather than the 120$^{\circ}$ ordered spin state. We thus expect to see peaks in the static spin structure factor with the same magnitude at the three degenerate stripy ordering wave vectors, namely, the $\mathbf{M}$ point and its equivalent in the Brillouin zone. The dominance of the static spin correlation at the $\mathbf{M}$ point over the $\mathbf{K}$ point, which corresponds to the 120$^{\circ}$ ordering pattern, is indeed seen in recent DMRG simulation on the TLHM\cite{Chen}. However, in a recent iDMRG simulation of the effective spin model\cite{Cookmeyer}, the author find that the spin correlation at the $\mathbf{K}$ point is stronger than that at the $\mathbf{M}$ point. A possible explanation for this contradictory result is that the parameter range used in Ref.[\onlinecite{Cookmeyer}] is too close to the phase boundary between the chiral spin liquid state and the 120$^{\circ}$ ordered state. 
 
In summary, we show that the chiral spin liquid state recently found from DMRG simulation of the TLHM in the intermediate coupling regime can be understood as a quantum disordered chiral spin ordered state - the tetrahedral spin state.  Such a spin state is stabilized by the four spin ring exchange coupling, which suppress the order-by-disorder effect toward a collinear stripy spin state and prefer an equal condensation at all the three degenerate stripy ordering wave vectors. We performed a Schwinger Boson mean field study of the $J_{1}-J_{2}-K_{4}$ model and find that such a chiral spin state prevails in a large portion of the phase diagram. However, we find that the Schwinger mean field theory can not describe correctly the topological property of such a state as a result of the lack of Boson rigidity in the mean field treatment. The mean field theory also overestimates the tendency of spinon condensation for the same reason. Both problems of the Schwinger Boson mean field theory call for a more sophisticated treatment in which the no double occupancy constraint is strictly enforced. Finally, we note that similar Bosonic RVB description can also be applied to other frustrated quantum antiferromagnets to address the origin of the chiral spin liquid state found in them, especially the Kagome Heisenberg antiferromagnet\cite{Gong2,He,Gong1}.

 \begin{acknowledgments}
We acknowledge the support from the grant NSFC 11674391, the Research Funds of Renmin University of China, and the grant National Basic Research Project
2016YFA0300504. 
\end{acknowledgments}
 
\section*{Appendix}
\setcounter{equation}{0}
\setcounter{subsection}{0}
\setcounter{figure}{0}
\renewcommand{\theequation}{A.\arabic{equation}}
\renewcommand{\thesubsection}{A.\arabic{subsection}}
\renewcommand{\thefigure}{A\arabic{figure}}

\subsection*{RVB ansatz for the three spin states studied in this work}

\begin{figure}
\includegraphics[width=7cm]{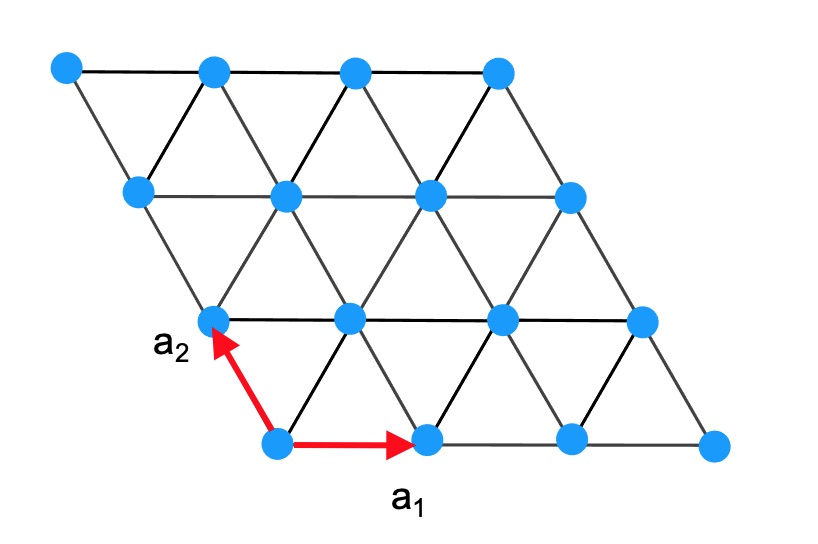}
\caption{The definition of the two lattice translation vector $\mathbf{a}_{1}$ and $\mathbf{a}_{2}$ of the triangular lattice. The momentum component $k_{1}$ and $k_{2}$ appearing below are defined as $k_{1}=\mathbf{k}\cdot \mathbf{a}_{1}$ and $k_{2}=\mathbf{k}\cdot \mathbf{a}_{2}$.}
\end{figure}

For the 120$^{\circ}$ spin state, the mean field ansatz can be made translational invariant after a suitable choice of gauge. The RVB parameter on the nearest neighboring bonds are given by
\begin{eqnarray}
A_{i,i+\mathbf{a}_{1}}&=&A_{i,i+\mathbf{a}_{2}}=-A_{i,i+\mathbf{a}_{1}+\mathbf{a}_{2}}=\mathcal{A}_{1}\nonumber\\
B_{i,i+\mathbf{a}_{1}}&=&B_{i,i+\mathbf{a}_{2}}=B_{i,i+\mathbf{a}_{1}+\mathbf{a}_{2}}=\mathcal{B}_{1},
\end{eqnarray}
in which $\mathcal{A}_{1}$ and $\mathcal{B}_{1}$ are two real parameters. The RVB parameter on the second neighboring bonds are given by
\begin{eqnarray}
A_{i,i+2\mathbf{a}_{1}+\mathbf{a}_{2}}&=&A_{i,i+2\mathbf{a}_{2}+\mathbf{a}_{1}}=A_{i,i+\mathbf{a}_{2}-\mathbf{a}_{1}}=0\nonumber\\
B_{i,i+2\mathbf{a}_{1}+\mathbf{a}_{2}}&=&B_{i,i+2\mathbf{a}_{2}+\mathbf{a}_{1}}=B_{i,i+\mathbf{a}_{2}-\mathbf{a}_{1}}=\mathcal{B}_{2},\nonumber\\
\end{eqnarray}
in which $\mathcal{B}_{2}$ is the third real parameter of the mean field Hamiltonian, which has the form of
\begin{equation}
H_{MF}=\sum_{\mathbf{k}}\left(\begin{array}{cc}b^{\dagger}_{\mathbf{k},\uparrow},b_{-\mathbf{k},\downarrow}\end{array}\right)\left(\begin{array}{cc} \xi_{\mathbf{k}} \ \ \Delta^{*}_{\mathbf{k}} \\ \Delta_{\mathbf{k}} \ \ \xi_{\mathbf{k}}\end{array}\right)\left(\begin{array}{c} b_{\mathbf{k},\uparrow} \\ b^{\dagger}_{-\mathbf{k},\downarrow}\end{array}\right).
\end{equation}
Here
\begin{eqnarray}
\xi_{\mathbf{k}}&=&\lambda+\mathcal{B}_{1}\gamma_{1}(\mathbf{k})+\mathcal{B}_{2}\gamma_{2}(\mathbf{k})\nonumber\\
\Delta_{\mathbf{k}}&=&\mathcal{A}_{1}\eta(\mathbf{k}),
\end{eqnarray}
with
\begin{eqnarray}
\gamma_{1}(\mathbf{k})&=&2[\cos k_{1}+\cos k_{2}+\cos (k_{1}+k_{2})]\nonumber\\
\gamma_{2}(\mathbf{k})&=&2[\cos( 2k_{1}+k_{2})+\cos(2k_{2}+k_{1})+\cos (k_{2}+k_{1})]\nonumber\\
\eta(\mathbf{k})& = & 2i[\sin k_{1}+\sin k_{2}-\sin (k_{1}+k_{2})].
\end{eqnarray}
Here $k_{1}=\mathbf{k}\cdot \mathbf{a}_{1}$, $k_{2}=\mathbf{k}\cdot \mathbf{a}_{2}$, with $\mathbf{a}_{1}$ and $ \mathbf{a}_{2}$ the two lattice vectors of the triangular lattice(see Fig.A1 for an illustration).

For the stripy spin state, the mean field ansatz can also be made translational invariant after a suitable choice of gauge. The RVB parameter on the nearest neighboring bonds are given by
\begin{eqnarray}
A_{i,i+\mathbf{a}_{1}}&=&0;\ A_{i,i+\mathbf{a}_{2}}= A_{i,i+\mathbf{a}_{1}+\mathbf{a}_{2}}=\mathcal{A}_{1}\nonumber\\
B_{i,i+\mathbf{a}_{1}}&=&\mathcal{B}_{1}; \ B_{i,i+\mathbf{a}_{2}}=B_{i,i+\mathbf{a}_{1}+\mathbf{a}_{2}}=0,
\end{eqnarray}
in which $\mathcal{A}_{1}$ and $\mathcal{B}_{1}$ are two real parameters. The RVB parameter on the second neighboring bonds are given by
\begin{eqnarray}
A_{i,i+2\mathbf{a}_{1}+\mathbf{a}_{2}}&=&A_{i,i+\mathbf{a}_{2}-\mathbf{a}_{1}}=\mathcal{A}_{2}; \ A_{i,i+2\mathbf{a}_{2}+\mathbf{a}_{1}}=0\nonumber\\
B_{i,i+2\mathbf{a}_{1}+\mathbf{a}_{2}}&=&B_{i,i+\mathbf{a}_{2}-\mathbf{a}_{1}}=0; \ B_{i,i+2\mathbf{a}_{2}+\mathbf{a}_{1}}=\mathcal{B}_{2},\nonumber\\
\end{eqnarray}
in which $\mathcal{B}_{2}$ is the third real parameter of the mean field Hamiltonian, which has the form of
\begin{equation}
H_{MF}=\sum_{\mathbf{k}}\left(\begin{array}{cc}b^{\dagger}_{\mathbf{k},\uparrow},b_{-\mathbf{k},\downarrow}\end{array}\right)\left(\begin{array}{cc} \xi_{\mathbf{k}} \ \ \Delta^{*}_{\mathbf{k}} \\ \Delta_{\mathbf{k}} \ \ \xi_{\mathbf{k}}\end{array}\right)\left(\begin{array}{c} b_{\mathbf{k},\uparrow} \\ b^{\dagger}_{-\mathbf{k},\downarrow}\end{array}\right).
\end{equation}
Here 
\begin{eqnarray}
\xi_{\mathbf{k}}&=&\lambda+\mathcal{B}_{1}\gamma_{1}(\mathbf{k})+\mathcal{B}_{2}\gamma_{2}(\mathbf{k})\nonumber\\
\Delta_{\mathbf{k}}&=&\mathcal{A}_{1}\eta_{1}(\mathbf{k})+\mathcal{A}_{2}\eta_{2}(\mathbf{k}),
\end{eqnarray}
with
\begin{eqnarray}
\gamma_{1}(\mathbf{k})&=&2\cos k_{1}\nonumber\\
\gamma_{2}(\mathbf{k})&=&2\cos( k_{1}+2k_{2})\nonumber\\
\eta_{1}(\mathbf{k})&=&2i[\sin k_{1}+\sin(k_{1}+ k_{2})]\nonumber\\
\eta_{2}(\mathbf{k})&=&2i[\sin( 2k_{1}+k_{2})+\sin(k_{2}-k_{1})].\nonumber\\
\end{eqnarray}

For the tetrahedral spin state, it is impossible to choose a gauge to make the mean field ansatz explicitly translational invariant. Thus we will simply set $\varphi_{i}=0$ in Eq.7. The RVB parameter of this mean field ansatz is given in Eq.11 of the main text. In the gauge we choose the unit cell of the mean field Hamiltonian contains four inequivalent sites. Such a mean field ansatz describes the so called weakly symmetric spin liquid state, which is invariant under space translation, proper space rotation and the combined operation of space inversion and time reversal(but not individually). According the naming scheme used in Ref.[\onlinecite{Messio}], the ansatz studied here belongs to the class with $\epsilon_{R_{6}}=1$, $\epsilon_{\sigma}=-1$, $p_{1}=1$, $k=1$ and $\phi_{B_{1}}=\pi/2$. We note that there is difference in the gauge used in this work and that used in Ref.[\onlinecite{Messio}]. The gauge transformation relating them is given by
\begin{equation}
\theta(i_{1},i_{2})=(-1)^{i_{2}}\times \mathrm{mod}(i_{1}-i_{2},4)\times\frac{\pi}{2},
\end{equation}
in which $\theta(i_{1},i_{2})$ denotes the phase change on site $\mathbf{r}_{i}=i_{1}\mathbf{a}_{1}+i_{2}\mathbf{a}_{2}$. The independent matrix element of the mean field Hamiltonian for the tetrahedral spin state is given by
\begin{eqnarray}
M_{1}(1,2)&=&\mathcal{B}_{1}(1+e^{-i2k_{1}})+\mathcal{B}_{2}(e^{i2k_{2}}+e^{-i2(k_{1}+k_{2})})\nonumber\\
M_{1}(1,3)&=&i\mathcal{B}_{1}(1+e^{-i2(k_{1}+k_{2})})+i\mathcal{B}_{2}(e^{-i2k_{1}}+e^{-i2k_{2}})\nonumber\\
M_{1}(1,4)&=&-\mathcal{B}_{1}(1+e^{-i2k_{2}})-\mathcal{B}_{2}(e^{i2k_{1}}+e^{-i2(k_{1}+k_{2})})\nonumber\\
M_{1}(2,3)&=&\mathcal{B}_{1}(1+e^{-i2k_{2}})+\mathcal{B}_{2}(e^{i2k_{1}}+e^{-i2(k_{1}+k_{2})})\nonumber\\
M_{1}(2,4)&=&-i\mathcal{B}_{1}(e^{i2k_{1}}+e^{-i2k_{2}})-i\mathcal{B}_{2}(1+e^{i2(k_{1}-k_{2})})\nonumber\\
M_{1}(3,4)&=&-i\mathcal{B}_{1}(1+e^{i2k_{1}})-i\mathcal{B}_{2}(e^{-i2k_{2}}+e^{i2(k_{1}+k_{2})})\nonumber\\
\end{eqnarray}
and 
\begin{eqnarray}
M_{3}(1,2)&=&\gamma\mathcal{A}_{1}(1+e^{-i2k_{1}})+\gamma\mathcal{A}_{2}(e^{i2k_{2}}+e^{-i2(k_{1}+k_{2})})\nonumber\\
M_{3}(1,3)&=&i\mathcal{A}_{1}(1+e^{-i(2k_{1}+k_{2})})+i\mathcal{A}_{2}(e^{-i2k_{1}}+e^{-i2k_{2}})\nonumber\\
M_{3}(1,4)&=&\gamma^{2}\mathcal{A}_{1}(1+e^{-i2k_{2}})+\gamma^{2}\mathcal{A}_{2}(e^{i2k_{1}}+e^{-i2(k_{1}+k_{2})})\nonumber\\
M_{3}(2,3)&=&\gamma^{2}\mathcal{A}_{1}(1+e^{-i2k_{2}})+\gamma^{2}\mathcal{A}_{2}(e^{i2k_{1}}+e^{-i2(k_{1}+k_{2})})\nonumber\\
M_{3}(2,4)&=&i\mathcal{A}_{1}(e^{i2k_{1}}+e^{-i2k_{2}})+i\mathcal{A}_{2}(1+e^{i2(k_{1}-k_{2})})\nonumber\\
M_{3}(3,4)&=&i\mathcal{A}_{1}(1+e^{i2k_{1}})+i\mathcal{A}_{2}(e^{-i2k_{2}}+e^{i2(k_{1}+k_{2})})\nonumber\\
\end{eqnarray}
The other matrix element of the Hamiltonian matrix can be found from the relation 
\begin{eqnarray}
M_{1}&=&M^{\dagger}_{1}\nonumber\\
M_{2}(\mathbf{k})&=&M^{T}_{1}(-\mathbf{k})\nonumber\\
M_{3}(\mathbf{k})&=&-M^{T}_{3}(-\mathbf{k}).
\end{eqnarray}

\end{document}